\documentstyle[art12,aaspp4,psfig,flushrt]{article}
\begin{document}

\title{Infrared Spectroscopy of GX 1+4/V2116 Oph: \\
    Evidence for a Fast Red Giant Wind?}

\author{Deepto~Chakrabarty,\altaffilmark{1} 
  Marten~H.~van~Kerkwijk,\altaffilmark{2} 
  and James~E.~Larkin\altaffilmark{3}}
\affil{\footnotesize Palomar Observatory, California Institute of
  Technology, Pasadena, CA 91125; \\
  deepto@space.mit.edu, mhvk@ast.cam.ac.uk, larkin@astro.ucla.edu}

\altaffiltext{1}{NASA Compton GRO Postdoctoral Fellow. Current
  address: Center for Space Research, Massachusetts Institute of
  Technology, Cambridge, MA 02139.}
\altaffiltext{2}{Current address: Institute of Astronomy, Madingley
  Road, Cambridge CB3 0HA, England, UK.}
\altaffiltext{3}{Current address: Department of Physics and Astronomy,
  University of California at Los Angeles, Los Angeles, CA 90095.}

\medskip
\centerline{\footnotesize\em Submitted 1997 December 15; accepted 1998
February 11} 
\medskip
\centerline{To appear in \sc The Astrophysical Journal Letters}

\begin{abstract}
We present infrared spectroscopy of the low-mass X-ray binary
GX 1+4/V2116 Oph.  This symbiotic binary consists of a 2-min
accretion-powered pulsar and an M5 III red giant.  A strong
\ion{He}{1} 1.083 $\mu$m emission line with a pronounced P Cygni 
profile was observed.  From the blue edge of this feature, we infer an
outflow velocity of $250\pm50$ km s$^{-1}$.  This is an order of
magnitude faster than a typical red giant wind, and we suggest that
radiation from the accretion disk or the neutron star may contribute
to the acceleration of the outflow.  We infer a wind mass 
loss rate of $\sim 10^{-6} M_\odot$ yr$^{-1}$.  Accretion from such a
strong stellar wind provides a plausible alternative to Roche lobe
overflow for supplying the accretion disk which powers the X-ray
source. The \ion{H}{1} Pa$\beta$ and \ion{He}{1} 1.083 $\mu$m emission
lines showed no evidence for the dramatic variability previously
reported in some optical lines, and no evidence for pulsations at the
2-min pulsar period. 
\end{abstract}


\clearpage

\section{INTRODUCTION}

Low-mass X-ray binaries (LMXBs) consist of a late-type star or
degenerate dwarf transferring matter to a neutron star or black hole
via an accretion disk.  The mass donor generally fills its
Roche lobe.  The optical emission in LMXBs is usually dominated by
reprocessing in the X-ray heated disk.  However, in those
systems with late-type donors and wide ($P_{\rm orb}\gtrsim 1$ d)
orbits, the infrared luminosity of the donor is comparable to that of
the disk, allowing direct study of the donor.  The infrared emission
is also less affected by interstellar absorption, which is generally a
significant hindrance in optical studies of LMXBs. 

The most luminous known LMXB mass donor is V2116 Oph, the M5 III giant
companion of the 2 min accretion-powered pulsar GX 1+4 (Glass \& Feast
1973; Davidsen, Malina, \& Bowyer 1977; Chakrabarty \& Roche 1997,
hereafter CR97).  The system is exceptional in a number of ways.  It
is a symbiotic binary (see Kenyon 1986 and Iben \& Tutukov 1996 for
reviews), the only one known with a confirmed neutron star companion.
Its binary period is unknown but must be $\gtrsim 100$ d, making it
the widest known LMXB by an order of magnitude.  The dense emission
line nebula surrounding the binary is powered by ultraviolet radiation
from an accretion disk.  Dramatic 100\% variability on a time scale of
minutes was observed in the optical emission fluxes of selected
\ion{He}{1} and \ion{Fe}{2} lines, while the continuum and all other
lines remained constant (CR97).  Coherent optical pulsations at the
2-min X-ray pulsar period are intermittently detected in the broadband
blue continuum (Jablonski et al. 1997).  GX 1+4 (=4U 1728--247) is one
of brightest hard X-ray sources in the Galactic center region.  X-ray
timing of the pulsar has revealed remarkable bimodal accretion torque
transitions between long intervals of steady, rapid spin-up and
equally rapid spin-down (Chakrabarty et al. 1997).  This may indicate
that the dipole magnetic field strength at the neutron star surface is
extraordinarily strong ($\sim 10^{14}$ G), or that the accretion disk
alternates between states of prograde and retrograde rotation and is
fed by the red giant's wind rather than by Roche lobe overflow
(Makishima et al. 1988; Chakrabarty et al. 1997; Nelson et al. 1997).

We report here on infrared observations of the GX 1+4/V2116 Oph
binary.  Besides characterization of the infrared spectrum of V2116
Oph, one of our primary goals was to exploit the bright infrared
emission from the star to search for the rapid emission line
variability reported by CR97 in the optical.  An unanticipated result
of our observations was the detection of a fast outflow from the M
giant, which has important implications in understanding the mass
transfer in this binary.

\section{OBSERVATIONS}

We observed V2116 Oph on 1995 May 10 using the new near-infrared
long-slit spectrometer (Larkin et al. 1996) at the $f/70$ Cassegrain
focus of the Palomar Observatory 5-m Hale telescope.  The two
available gratings yield wavelength resolutions
$R\equiv\lambda/\Delta\lambda\approx 1000$ and $R\approx 4000$, when
used with a 0.7 arcsec slit width.  We made two types of observations:
relatively long exposures (several pulsar periods) to get time-averaged
spectra of the object, and rapid time sequences to look for time
variability in selected emission lines.

For the long exposures, five $R\approx 1000$ spectra were taken.
These included four 1-min observations at different grating angles to
cover the entire $K$ bandpass (2.0--2.4 $\mu$m), and a 2-min
observation within the $J$-band, centered on the \ion{H}{1} 1.2818
$\mu$m (Paschen $\beta$) line.  Also, a high resolution ($R\approx
4000$) spectrum of the \ion{He}{1} 1.083 $\mu$m line was obtained with
a total integration time of 6 min.

All of the observations were made in pairs using a 0.75$\times$40
arcsec slit.  The first exposure of a pair was taken with the target
10 arcsec from one end of the slit, and the target was then moved 20
arcsec along the slit for the second exposure.  The spectra were
reduced by subtracting image pairs and then dividing by a
sky-subtracted standard star frame.  Bad pixels were removed by linear
interpolation.  To remove a slight curvature along the spatial axis, a
star was moved in 5 arcsec intervals along the slit and its position
was fit with a third-order polynomial.  For all our observations, the
seeing was $\lesssim 1$ arcsec at 2 $\mu$m, and the sky was
photometric.  A nearby star was simultaneously observed with an offset
guider which controlled a tip-tilt secondary.  With this guiding
system in place, tracking errors ($\lesssim$0.1 arcsec) were
insignificant compared to the seeing.

Wavelength calibration was achieved by fitting atmospheric OH airglow
emission lines with a quadratic polynomial.  We obtained the OH line
wavelengths from Oliva \& Origlia (1992) and Maihara et al. (1993).  A
conservative estimate of the absolute wavelength calibration
uncertainty is roughly one pixel, corresponding to 70 km s$^{-1}$ for
$R\approx 1000$ and 15 km s$^{-1}$ for $R\approx 4000$.  This includes
the systematic uncertainty due to curvature of the spectral lines and
the line fitting procedure.  The observed wavelengths of V2116 Oph
were corrected for the Earth's motion with respect to the solar system
barycenter.  Air wavelengths are quoted for all spectral features
discussed in this paper.  

The flux calibration was performed in the following way.  For the
relative calibration, we used observations of the G3 IV standard star
HR 6441, for which we assumed a blackbody temperature of 5725~K.  As we
only observed through a narrow slit, absolute calibration using HR
6441 was not possible.  Instead, we normalized our spectra to the
typical infrared photometric magnitudes of V2116 Oph, which have been
fairly constant over the 1990s except during the bright X-ray flare of
1993 September (CR97). 

The time variability of the Pa$\beta$ and \ion{He}{1} 1.083 $\mu$m
lines was investigated by obtaining 6 sequences of time-resolved
spectra of each line.  Each sequence consisted of 8 exposures with an
integration time of 8 s.  Due to overhead in reading out and storing
the data, exposures were actually separated by 14.43 s, and the total
duration of a sequence was 115.455 s (close to the 2 min pulse period
of GX~1+4).  Between sequences (i.e., every 2 min), the object was
moved to the opposite end of the slit, as described above.   The
sequences were reduced in pairs, with the average spectrum of one
sequence being used for the sky subtraction of the individual spectra
in the other.  For each line, a total of 48 exposures (each 8 seconds
long) was obtained, covering a total time span of about 1000 seconds.
The resolution was $R\approx 1000$ for the Pa$\beta$ spectra and
$R\approx 4000$ for the \ion{He}{1} 1.083 $\mu$m spectra.  

\section{RESULTS}

Figure~1 shows our combined $K$-band spectrum of V2116 Oph.  In
addition to strong \ion{H}{1} 2.1655 $\mu$m (Brackett $\gamma$)
emission and absorption band features due to $^{12}$CO and $^{13}$CO,
there are several weaker absorption features due to neutral metal
species.  The \ion{He}{1} 2.06 $\mu$m line may also be detected in
emission, but telluric contamination shortward of 2.1 $\mu$m prevents
a secure identification.  The spectrum longward of 2.1 $\mu$m is
similar to one obtained contemporaneously from UKIRT (at somewhat
lower resolution) by Bandyopadhyay et al. (1997) on 1995 June 29.
Except for the Br$\gamma$ emission line, both observations are
consistent with the spectrum expected from an isolated late-M giant
(Kleinmann \& Hall 1986).  Comparing our CO absorption band
measurements with the synthetic spectra of Lazaro et al. (1991), we
estimate that the [C/H]$\approx 0.0$ and $^{12}$C/$^{13}$C$\approx 5$.
The latter quantity is consistent with an independent estimate of
$^{12}$C/$^{13}$C$\approx 8$ by Bandyopadhyay et al. (1997).  By
comparison, both single M giants and M giants in symbiotics generally
have carbon relative and isotopic abundances somewhat below the solar
value (Lazaro et al. 1991; Schild, Boyle, \& Schmid 1992).

Time-resolved emission line photometry of the Pa$\beta$ and
\ion{He}{1} lines showed no evidence for the dramatic 100\%
variability on a 10 min time scale previously reported for some
optical \ion{He}{1} and \ion{Fe}{2} emission lines (CR97).  The two
lines showed fractional root-mean-squared variations of about 15\%,
which is consistent with Poisson counting statistics.  We folded the
IR spectrophotometric data at the X-ray pulse period of the neutron
star (122.169 s on 1995 May 10.44; Chakrabarty et al. 1997) to look
for coherent pulsations.  No pulsations were detected in either the
emission lines or the continuum.  The 2$\sigma$ upper limits on the
pulsed fractions were $<5$\% for the \ion{He}{1} line and $<13$\% for
the adjoining continuum, and $<9$\% for the Pa$\beta$ line and $<5$\%
for the adjoining continuum.  For comparison, pulsation searches in
optical emission lines set upper limits of $<1.7$\% (3$\sigma$) in
H$\alpha$ (Krzeminski \& Priedhorsky 1978) and $<3$\% in \ion{O}{1}
8446\,\AA\ (Deutsch, Margon, \& Bland-Hawthorne 1997).  However,
optical pulsations from V2116 Oph have been intermittently detected in
the broadband blue continuum with 2\% amplitude (Jablonski et
al. 1997).  Since the pulsed fraction of the X-ray emission is known
to be high ($\gtrsim 50$\%), the absence of pulsations in the emission
lines is consistent with the conclusion of CR97 that the emission
lines are powered by ultraviolet radiation from the accretion disk
rather than X-rays from the accreting pulsar.  The cause of the rapid
optical line variability reported by CR97 remains a puzzle.

We attempted to use the observed line features to measure the radial
velocity of V2116 Oph.  The Pa$\beta$ line gave a velocity of
$-255\pm70$ km s$^{-1}$, and the Br$\gamma$ line gave a velocity of
$-190\pm70$ km s$^{-1}$.  Velocity measurements with the weak
absorption features in the $K$-band spectrum were difficult, because
the lines are inherently asymmetric (Kleinmann \& Hall 1986) and were
measured at low resolution; however, the \ion{Na}{1} and \ion{Fe}{1}
features gave velocities ranging from $-120$ km s$^{-1}$ to $-370$ km
s$^{-1}$.  In principle, the strong CO band features can provide an
excellent velocity measurement.  However, we did not attempt this
since our wavelength calibration redward of about 2.3 $\mu$m is
uncertain, due to the lack of OH airglow lines in that region.  For
comparison, the H$\alpha$ 6563~\AA\ emission line has a velocity of
$\approx -150$ km s$^{-1}$ (CR97).

Figure 2 shows the line profile for the \ion{He}{1} 1.0830 $\mu$m
emission feature.  Weak emission (equivalent width $-W_\lambda\sim 30$
m\AA) in this line has been observed in several M giants, three of
which are in binaries (O'Brien \& Lambert 1986).  However, in V2116
Oph this line is three orders of magnitude stronger ($-W_\lambda=$31
\AA) and is probably powered by photoionization of the red giant's
wind by the accretion disk.  A pronounced P Cygni profile is evident in
Figure 2.  The blue edge of the absorption core has a velocity of
$-473\pm15$ km s$^{-1}$, and the red edge of the emission line has a
velocity of $-5\pm15$ km s$^{-1}$.  P Cygni profiles are caused by an
expanding shell around a star (e.g., Mihalas 1978).  The absorption
core is due to material in front of the stellar disk, while the
emission line is due to material on either side of the star.  The
maximum outflow velocity of the material is given by the velocity
difference of the blue edge of the absorption core and the rest frame
of the star.  (The red edge of the emission line cannot be used, since
the star occults the material moving at maximum velocity.)  We do not
know the velocity of V2116 Oph securely.  However, adopting the
average of the Pa$\beta$ and Br$\gamma$ velocities, $-220\pm50$ km
s$^{-1}$, we infer an outflow velocity of $250\pm50$ km s$^{-1}$ from
V2116 Oph.  If we instead adopt the H$\alpha$ velocity of $-150$ km
s$^{-1}$ for V2116 Oph, then the inferred outflow velocity is $\approx
320$ km s$^{-1}$.  No clear evidence for a P Cygni profile was found
in the hydrogen Paschen or Brackett lines.

Late M giants generally have relatively strong \ion{Si}{1} 1.0827
$\mu$m and \ion{Ti}{1} 1.0828 $\mu$m absorption lines (O'Brien  \&
Lambert 1986).  We now consider whether these features could distort
the intrinsic P Cygni profile of the \ion{He}{1} 1.083 $\mu$m line.  
If we shift these features by our assumed radial velocity of 
$-220$ km s$^{-1}$ for V2116 Oph, then they would overlap the red edge
of the P Cygni absorption core but would leave the blue edge of the
absorption core unaffected.  Also, the \ion{Si}{1} and \ion{Ti}{1}
features have a combined equivalent width of $W_\lambda\approx 0.33$
\AA\ and a central depth of 25\% relative to the continuum in late M
giants (O'Brien \& Lambert 1986), compared to $W_\lambda=3.3$ \AA\ and a 
central depth of 80\% for our P Cygni absorption core.  Thus, we are
confident that our outflow velocity is unaffected by the blended lines.  

%\clearpage
\section{DISCUSSION}

It is widely believed that red giants have slow winds (10--30 km
s$^{-1}$; e.g., Reimers 1981, Dupree 1986).  Since these winds are
much slower than the photospheric escape velocity $v_{\rm
esc}=(2GM_g/R_g)^{1/2}$, the wind acceleration mechanism must be
operating in the extended atmosphere of the giant.  However, the
outflow velocity we observe from V2116 Oph is an order of magnitude
larger than a typical red giant wind velocity.  Most existing
measurements of wind velocities in late type giants were made in large
stars on the asymptotic giant branch (AGB), with $R_g \gtrsim 100
R_\odot$.  Since we expect the wind velocity to scale roughly as
$R_g^{-1/2}$, a smaller star might have a somewhat faster wind.
Indeed, a fast ($\sim 100$ km s$^{-1}$) wind was observed from the
metal-poor G9.5 III Fe-2 star HD 6833, which has roughly solar mass
and a radius of 30--45 $R_\odot$ (Dupree, Sasselov, \& Lester 1992).
Based on luminosity and extinction arguments, CR97 concluded that
V2116 Oph is probably near the tip of the first-ascent red giant
branch, with a radius $R_g\lesssim 100 R_\odot$.  However, this
argument seems unable to account for an order of magnitude increase in
the wind velocity, since V2116 Oph is certainly not a factor of 100
smaller in radius than an AGB star.

It is possible that the observed outflow is due to a wind from the
accretion disk rather than the red giant.  The X-ray timing behavior
of the pulsar (Chakrabarty et al. 1997), the optical emission line
spectrum (CR97), and the optical pulsations in the broadband blue
continuum (Jablonski et al. 1997) all point to the presence of an
accretion disk in this binary.  However, our observed velocity seems
too {\em small} for mass loss from the disk, since accretion disk
winds are radiatively driven and supersonic (see, e.g., Warner 1995
and references therein).

Although isolated late-M giants may be unable to produce fast winds,
it is possible that the energy for accelerating the outflow is
provided by the hot component in a symbiotic binary.  In the case of
V2116 Oph, either the ultraviolet radiation from the neutron star's
accretion disk or X-rays from the neutron star itself certainly
possess sufficient energy to power a fast wind.  While direct
radiative acceleration of the wind is inconsistent with our data,
illumination of the red giant's atmosphere might lead indirectly to
enhanced acceleration of an outflow.  It is interesting to
note that a strong \ion{He}{1} 1.083 $\mu$m P Cygni profile indicative
of a fast ($\approx 150$ km s$^{-1}$) wind was also observed from the
symbiotic star CI Cyg, an M5 II giant (Bensammar et al. 1988).  The
cool component in this eclipsing symbiotic binary is an AGB star,
while the hot component is thought to be a $\sim 0.5 M_\odot$ main
sequence star with an accretion disk (Kenyon 1986; Kenyon et
al. 1991).  The observed wind is probably from the cool component,
however, as it is only observed during eclipse.  As in V2116 Oph, the
hydrogen Brackett and Paschen emission lines in CI Cyg did show P Cygni
profiles.  This is not particularly surprising, though, since the
metastability of the lower state of the \ion{He}{1} 1.083 $\mu$m
transition makes it a particularly sensitive probe of bulk mass motion
in the atmospheres of cool stars (O'Brien \& Lambert 1986; Lambert
1987; Dupree et al. 1992). 

Whatever the acceleration mechanism, a fast wind has important
consequences in the GX 1+4/V2116 Oph  binary.  CR97
pointed out that V2116 Oph probably does not fill its
Roche lobe, since there is no evidence for the highly super-Eddington
mass transfer that would then be expected from such a large star.
Instead, they prefer the suggestion of Makishima et al. (1988) that
the neutron star is accreting from the red giant wind.  Interpreting
the marginal detection of 6 cm radio continuum emission from V2116 Oph
(Marti et al. 1997) as thermal bremsstrahlung due to ionization of the
red giant wind by the accretion disk and/or the X-ray pulsar,
CR97 inferred a wind mass-loss rate of
\begin{equation}
-\dot M_w \approx 1\times 10^{-7} M_\odot \mbox{\rm\,yr$^{-1}$}
    \left(\frac{v_w}{\mbox{\rm 10\ km s$^{-1}$}}\right)
    \left(\frac{D}{\rm 10\ kpc}\right)^{3/2},
\end{equation}
where $v_w$ is the wind velocity and $D$ is the distance to the
source.  For a 10 km s$^{-1}$ wind, at least 1\% of the wind would
have to be accreted by the pulsar in order to explain the $\gtrsim
10^{37}$ erg s$^{-1}$ X-ray luminosity observed from GX 1+4.  However,
the required capture fraction drops to $\sim 0.1$\% if the wind
velocity is an order of magnitude higher.   Low capture efficiencies
are typical of wind accretion in X-ray binaries (see Frank, King, \&
Raine 1992). 

We can use the P Cygni profile to make an independent estimate of the
wind mass loss from V2116 Oph.  Naively assuming that the outflow
consists of ionized hydrogen, and adopting the characteristic density
and scale size of the symbiotic nebula enshrouding the giant (CR97),
the continuity equation yields
\begin{equation}
-\dot M_w \approx 2\times 10^{-6} M_\odot \mbox{\rm\,yr$^{-1}$}
    \left(\frac{n_e}{10^9 \mbox{\rm\,cm$^{-3}$}}\right)
    \left(\frac{v_w}{250 \mbox{\rm\,km s$^{-1}$}}\right)
    \left(\frac{r_{\rm out}}{1 \mbox{\rm\,AU}}\right)^2 ,
\end{equation}
where $n_e$ is the electron density and $r_{\rm out}$ is the effective
radius of the outflow.  This is roughly consistent with the rate
inferred from the radio flux in equation (1), assuming $v_w=250$
km~s$^{-1}$.  With such a high mass loss rate from the red giant, only
a very small fraction of the wind need be accreted by the pulsar to
explain the observed X-ray luminosity. Thus, it seems plausible that
the accretion disk in GX 1+4 is supplied by the red giant's wind
rather than by Roche lobe overflow.  If this is the case, then V2116
Oph is the only known LMXB donor which does not fill its Roche lobe.

\acknowledgements{We thank an anonymous referee for a careful reading
of our manuscript and a number of helpful suggestions.  It is also a
pleasure to thank Keith Matthews for his assistance in making these
observations, especially the timing measurements.  We further thank
Reba Bandyopadhyay, Andrea Dupree, Icko Iben, Mike Jura, and Ron
Webbink for useful discussions; and Tony Lynas-Gray, Carlos Lazaro,
and Robin Clegg for sharing their synthetic CO band spectra for M
giants with us.  D.C. was supported by a NASA Compton GRO Postdoctoral
Fellowship at MIT under grant NAG 5-3109.  M.H.v.K. was supported by a
NASA Hubble Postdoctoral Fellowship at Caltech.}

%\vspace{1.3in}
%\vspace{1in}

\centerline{FIGURE CAPTIONS}

\noindent
FIGURE 1: Medium-resolution $K$-band spectrum of V2116 Oph.  Except for
the hydrogen Br$\gamma$ emission line, the spectrum is typical of an
isolated M5 III giant.   The \ion{He}{1} 2.06 $\mu$m line may be
detected in emission, but telluric contamination shortward of 2.1
$\mu$m prevents a secure identification.

\bigskip
\noindent
FIGURE 2: High-resolution P Cygni profile of the \ion{He}{1} 1.083
$\mu$m emission line in V2116 Oph.  Relative to the laboratory air
wavelength of 1.08303 $\mu$m, the blue edge of the absorption core has
a velocity of $-473\pm15$ km s$^{-1}$, and the red edge of the
emission line has a velocity of $-5\pm15$ km s$^{-1}$.  Adopting
$-220\pm50$ km s$^{-1}$ (the average of the Pa$\beta$ and Br$\gamma$
velocities) as the velocity of the star, we infer a wind speed of
$250\pm50$ km s$^{-1}$. 

\clearpage
\thispagestyle{empty}
\begin{figure}
\vspace{-2in}
\centerline{Figure 1}
\vspace{1in}
\centerline{\psfig{file=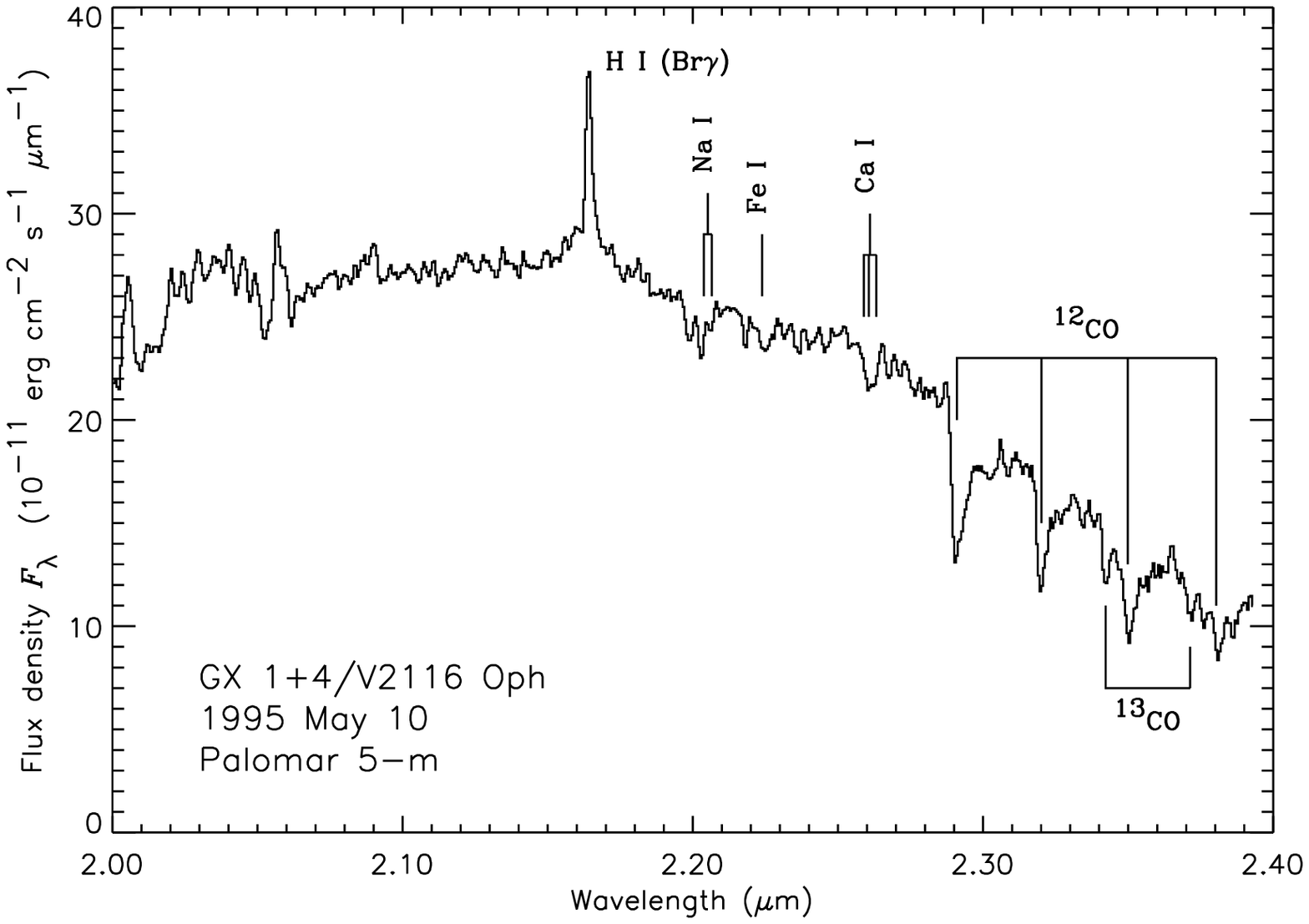}}
% ~/GX1+4/ir/makekband.pro
\end{figure}

\clearpage
\thispagestyle{empty}
\begin{figure}
\vspace{-0.5in}
\centerline{Figure 2}
\vspace{2in}
\centerline{\psfig{file=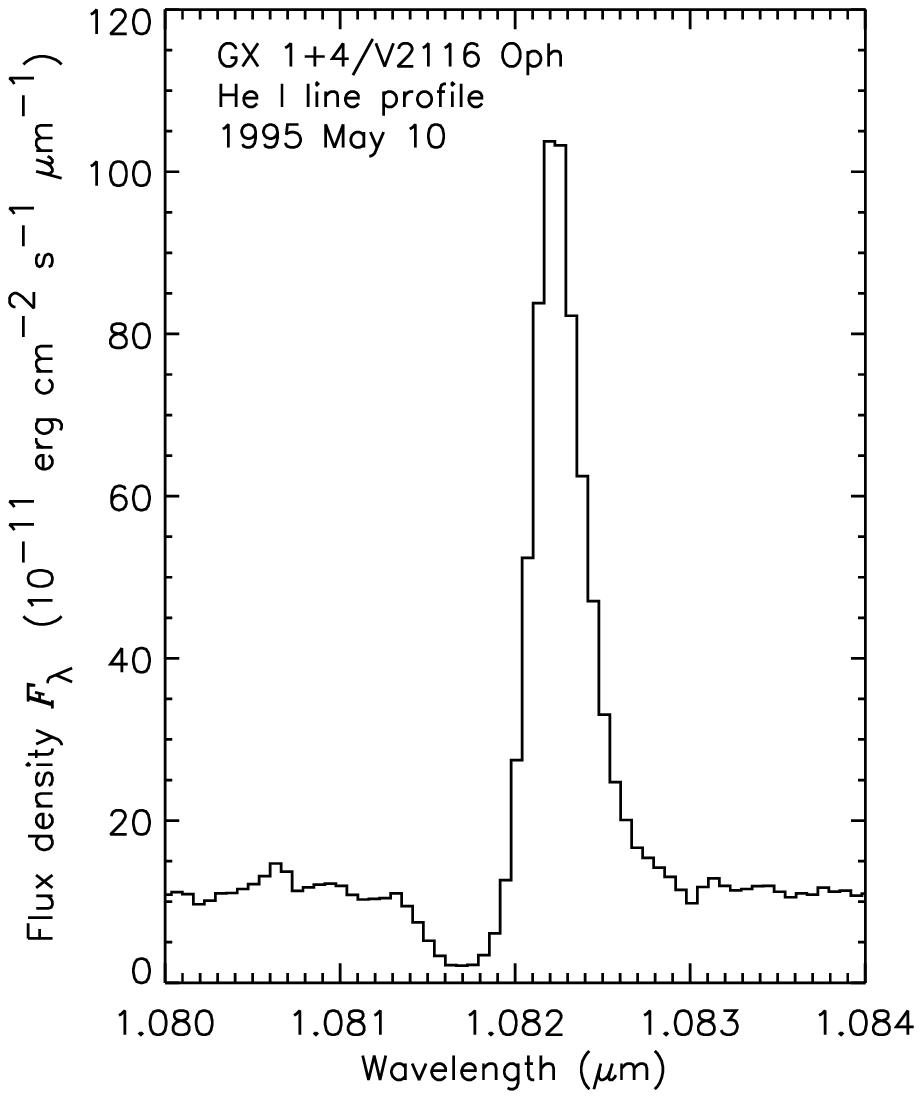}}
% ~/GX1+4/ir/make108.pro
\end{figure}

\end{document}